\documentstyle[aps,psfig,prl,multicol]{revtex}
\newlength{\FigWidth}
\begin{document}
\draft
\title{
Experimental observation of the mobility edge in a waveguide with correlated disorder}
\author{
U.~Kuhl$^1$, F.M.~Izrailev$^2$,  A. A.~Krokhin$^2$, and H.-J.~St\"ockmann$^1$}

\address{
$^1$ Fachbereich Physik, Universit\"at Marburg, Renthof 5,
D-35032 Marburg, Germany \\
$^2$Instituto de F\'{\i}sica, Universidad Aut\'onoma de Puebla, Apartado Postal J-48,
Puebla, 72570 M\'exico
}

\date{\today}
\maketitle

\begin{abstract}
The tight-binding model with correlated disorder introduced by
Izrailev and Krokhin [PRL {\bf 82},  4062  (1999)] has been
extended to the Kronig-Penney model. The results of the
calculations have been compared with microwave transmission
spectra through a single-mode waveguide with inserted correlated scatterers.
All predicted bands and mobility edges
have been found in the experiment, thus demonstrating that any
wanted combination of transparent and non-transparent frequency intervals
can be realized experimentally by introducing appropriate correlations between
scatterers.\end{abstract}

\pacs{PACS numbers: 72.15.Rn, 72.20.Ee, 73.20.Jc}

\begin{multicols}{2}

Starting from the pioneering paper by Anderson \cite{and58}, a lot
of progress has been achieved in the theoretical study of 1D
tight-binding models. This model includes a wide range of
different physical situations lying in between two limit cases:
ideal periodic lattices where all states are extended, and
completely random lattices where any state is exponentially
localized. Specific interest has been paid to the so-called
pseudo-random (or deterministic aperiodic) potentials which
demonstrate either localization or delocalization, depending on
their parameters \cite{bel82,gri88,sar88}. A widely used model is
described by the Harper equation with the site potential
$V_n=\epsilon \cos(2\pi\alpha n)$. For $\alpha$ irrational, the
incommensurability of the potential gives rise to a
localization-delocalization transition (for all states) when the
amplitude $\epsilon$ passes through the critical value
$\epsilon_{\rm cr}=2$, see e.g., Ref. \cite{gei95}. For fixed
$\epsilon$ the energy spectrum of the Harper equation exhibits the
famous Hofstadter butterfly \cite{hof76} when $\alpha$ scans the
interval $[0,1]$. This rather exotic spectrum was recently
observed experimentally \cite{kuh98b} by making use of the
equivalence of the Harper equation and the wave equation in a
single-mode electromagnetic waveguide with point-like scatterers.

For a long time a coexistence of localized and extended states in
the spectrum of eigenenergies of 1D random potentials was
considered to be impossible. However, it was shown in
Refs.~\cite{dun90,phi91} that a {\it discrete} set of delocalized states appears if
short-range correlations are introduced in a random potential. This is done by
repeating twice each value of site potential (dimer model). Recently discrete extended states have
been observed in the experiment with
GaAs-AlGaAs random superlattices \cite{bel99}.

A general case of 1D potential in tight-binding
approximation with arbitrary correlations was considered in Ref.
\cite{izr99}. A direct relation between the pair correlation function and the
localization length has been derived. This relation shows that the
mobility edge does exist in 1D geometry. A few examples of potentials with correlated disorder
were given. All these potentials necessarily contain the
long-range correlations which thus give rise to a {\it continuum} set of
delocalized states and to mobility edge.

\begin{figure}
\psfig{file=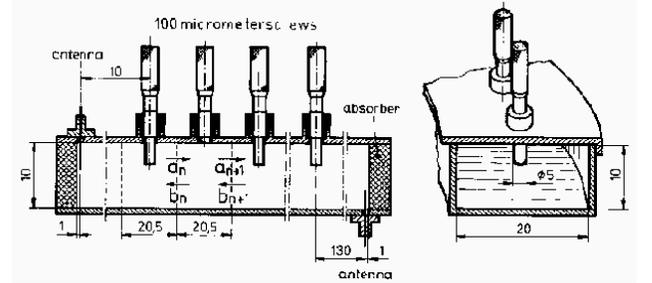,width=\FigWidth} \narrowtext
\caption{\label{fig:setup} Experimental set-up. All dimensions are
given in mm.}
\end{figure}

In this Letter we exploit the analogy between the propagation of
quantum particle and electromagnetic wave, in order to
demonstrate experimentally the existence of the mobility edge in
1D geometry. The experimental set-up is shown in
Fig.~\ref{fig:setup}. It is the same as has been already used
in the microwave realization of the Hofstadter butterfly
\cite{kuh98b}. From the top of a waveguide of total length of
2.15~m 100 micrometer screws can be turned in. By varying the
lengths of the micrometer screws different scattering arrangements
can be realized. Complete information about the scattering matrix of a given arrangement
is obtained via two antennas at the ends of the waveguide
using a Wiltron 360B network analyzer.

If the screws are approximated by delta scatterers, the
propagation of a single mode through the waveguide is described by the
wave equation for the Kronig-Penney model,
\begin{equation}
\label{base} \psi ^{\prime \prime }(z)+E\psi (z)=\sum_{n=-\infty
}^\infty EU_n\psi (z_n)\delta (z-nd).
\end{equation}
Here the wave function $\psi$ is associated with electric field of the $TE$-mode,
and the energy is given by $E=k^2$, where $k$ is the wavenumber.
We write Eq.~(\ref{base}) in the discrete form \cite{lif88},
\begin{equation}
\psi_{n+1}+\psi _{n-1}=
\left[ 2\cos (kd)+U_nkd \sin(kd)\right] \psi_n,
\label{discrete}
\end{equation}
where $\psi_n\equiv \psi(z_n)$. The potential
strength $U_n=\epsilon +\epsilon _n$ is split into two
parts, its mean value $\epsilon =\langle U_n \rangle$ and fluctuations
$\epsilon_n$. Our treatment is based on the approach \cite{izr95}
which allows one to express the quantum model
(\ref{base}) in terms of the classical two-dimensional Hamiltonian
map,
\begin{equation}
\label{map}
\begin{array}{cc}
p_{n+1}=(p_n\,+A_n\,x_n)\cos \mu\,\,-\,\,x_n\sin \mu, &  \\
x_{n+1}=(p_n\,+A_n\,x_n)\sin \mu \,\,+\,\,x_n\cos \mu, &
\end{array}
\end{equation}
where $\,x_n=\psi _n$ is the position and $p_n$ is the conjugate canonical
momentum. This map describes the
behavior of a linear oscillator subjected to linear periodic delta-kicks.
The amplitude $A_n$ of these kicks is defined as
\begin{equation}
A_n = \frac {k \epsilon_n \sin(kd)}{\sin \mu},
\label{kick}
\end{equation}
and the phase shift $\mu$
between two kicks is given by the
dispersion relation for the Kronig-Penney model
\begin{equation}
\label{shift}
2 \cos\mu =2\cos(kd) + kd\epsilon \sin(kd),\,\ 0\leq \mu \leq \pi\,.
\end{equation}
The parameter $\mu$ plays the role of the Bloch number and the width of the
Bloch band is defined by $\epsilon$.
In this approach localized
quantum states correspond to trajectories which are unbounded in the
classical phase space $(p,x)$ when $n\rightarrow \infty $. Contrary,
extended states are represented by bounded trajectories.

It is convenient to introduce the action-angle variables
$(r,\theta )$ according to the standard relations, $x=r\,\sin
\,\theta ,\,p=r\,\cos \,\theta$. Then the inverse localization
length can be defined as
\begin{equation}
\label{locdef}l^{-1}(E)=\lim _{N\rightarrow \infty }\,\frac
1N\sum_{n=1}^N\ln \left( \frac{r_{n+1}}{r_n}\right) \,\,,
\end{equation}
where
\begin{equation}
\label{Dn}
\frac{r_{n+1}}{r_n}=\sqrt{1+A_n\sin({2\theta _n)}+A_n^2\sin^2\theta _n}.
\end{equation}
This definition coincides with the standard one
$\l^{-1} =\langle \ln |\psi _{n+1}/\psi_n|\rangle $ \cite{lif88}
inside the allowed energy bands \cite{izr98}.
Here the brackets stand for the average over $n$.

This Hamiltonian map approach turns out to be very effective in the
study of completely disordered potentials \cite{izr98} as well as potentials
with correlated disorder \cite{izr99,izr95}. In particular, in Ref. \cite{izr99}
the expression for the localization length for the tight-binding model
with {\it any} kind of correlations in the potential has been obtained.
Since the relation (\ref{discrete}) has the form of the
tight-binding model, one can use the results of Ref. \cite{izr99}.
Then the inverse localization length for the Kronig-Penney model reads,
\begin{eqnarray}
\label{locfin}
l^{-1}(E)&=&
k^2 \frac{\langle\epsilon_n^2\rangle}{8}\frac{\sin ^2(kd)}{\sin ^2\mu }\varphi (\mu ), \\
\label{locfinvar}
\varphi(\mu)&=&
1+2\sum\limits_{m=1}^\infty \xi_m\,\cos \,(2\mu \,m) .
\end{eqnarray}
where $\xi_m=\langle \epsilon_{n+m}\epsilon_n\rangle/\langle\epsilon_n^2\rangle$
is the dimensionless binary correlator.

Relation (\ref{locfin}) is a starting point to obtain the conditions under which the
mobility edges exist for 1D random potentials.
The function $\varphi(\mu)$ in Eq.~(\ref{locfinvar}) is defined through
its Fourier coefficients $\xi_m$.
Then for a given dependence $\varphi(\mu)$ the correlators
can be calculated via
\begin{equation}
\label{ksi}\xi_m=\frac 2\pi \int\limits_0^{\pi /2}\varphi \,(\mu )\,\cos
(2m\mu )\,d\mu.
\end{equation}
Now the problem is reduced to the construction of potentials having
given correlators. Leaving mathematical details for a complete publication\cite{unpub},
we give here the final formula for set of random potentials with appropriate binary
correlation function\cite{ThanksSok}:

\begin{eqnarray}
\label{alg}
\epsilon _n &=& \sqrt{\langle\epsilon_n^2\rangle}
\sum\limits_{m=-\infty}^{+\infty} \beta_m Z_{n+m}, \\
\label{beta}
\beta_m   &=& \frac 2\pi \int_0^{\pi /2}\sqrt{\varphi (\mu )} \cos(2m\mu )d\mu.
\end{eqnarray}
Here $Z_n$ are random numbers with mean zero and variance one.
It is easy to check that the dimensionless correlator of the site potential
Eq.~(\ref{alg}) coincides with the Fourier coefficient (\ref{ksi}).
Eqs.~(\ref{locfin})-(\ref{beta}) give an explicit solution of the
inverse problem, since they reconstruct random potentials from
the dependence $l(E)$.
The existence of the sharp mobility edges means that the function
$l^{-1}(E)$ has only finite
(small) number $\nu$ of derivatives at the corresponding energies.
Then, the Fourier coefficients (\ref{ksi}) decay slowly, $\xi_m \sim
m^{-(\nu+1)}$ for $m \gg 1$.
This means that mobility edges are due to the {\it long-range
correlations} in the random potential. Numerical data \cite{izr99}
for the step-function dependence ($\nu=0$) of $\varphi(\mu)$
show that the localization length, indeed, reveals sharp
mobility edges. Mobility edges obtained
for self-affine potentials \cite{mou98} are also due to the long-wavelength component
of correlation function.

Let us now construct a random potential which results in the following
function $\varphi(\mu)$ (we chose this function to be
symmetrical with respect to the point $\mu=\pi/2$),

\begin{equation}
\varphi(\mu)=\left\{
\begin{array}{cc}
C_0^2, & 0<\mu_1<\mu<\mu_2 <\pi/2 \\
0, & \mu<\mu_1; \, \pi/2>\mu>\mu_2 \,\,.
\end{array}
\right. .
\label{l1}
\end{equation}
\noindent
Here $C_0^2=\pi/
2(\mu_2-\mu_1)$ is the normalization constant
obtained from the condition $\xi_0=1$.  This dependence exhibits
four sharp ($\nu=0$)
mobility edges in the first
allowed zone.
Their positions are given by two pairs of roots of Eq. (\ref{shift})
with $\mu=\mu_1$ and $\mu=\mu_2$. Using Eq.~(\ref{beta}) one obtains
$\beta_0=2C_0(\mu_2-\mu_1)/\pi$ and
\begin{eqnarray}
\label{beta1}
\beta_m=
\frac {C_0}{\pi m} \left\{\sin(2m\mu_2) - \sin(2m\mu_1) \right \}.
\end{eqnarray}
\begin{figure}
\psfig{file=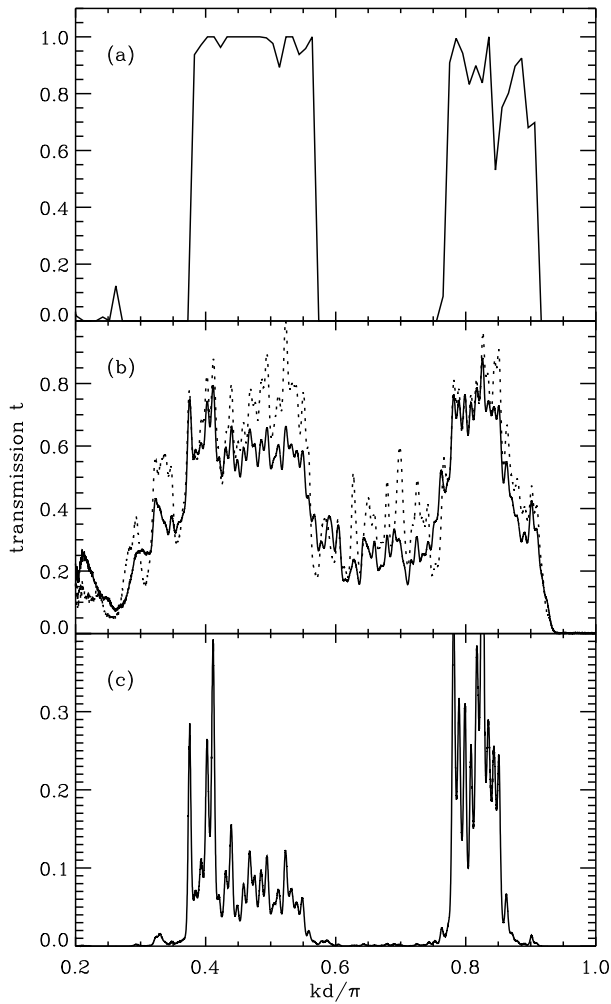,width=\FigWidth} \narrowtext
\caption{\label{fig:band1} Transmission through the 1D random
sequence with correlation governed by Eq.~(\ref{l1}): (a)
numerical results for $N=10^4$ scatterers,
(b) microwave transmission through an array of
$N=100$ scatterers (dotted line) and average over five different
measurements (solid line), (c) microwave transmission through an
array of $N=500$ scatterers obtained by multiplicating the
transfer matrices of five individual measurements.
}
\end{figure}

Experimentally it is difficult to measure the localization length directly.
The accessible quantity is the transmission coefficient for a finite sample.
Similar to the localization length, the transmission coefficient $t_N$
can be expressed in terms of the classical map (\ref{map}) \cite{izr95},
\begin{equation}
t_N={\frac{4 }{2 + r_{1N}^2+r_{2N}^2}}.
\label{trans}
\end{equation}
Here $r_{1N}$ and $r_{2N}$ are the radii of the trajectories at time $N$,
starting at a radius $r_0=1$ and angles $\theta_0=0$ and $\pi/2$, respectively.
This geometrical
interpretation of the transmission coefficient is very useful
for understanding its generic properties as well as for numerical simulations.

\begin{figure}
\psfig{file=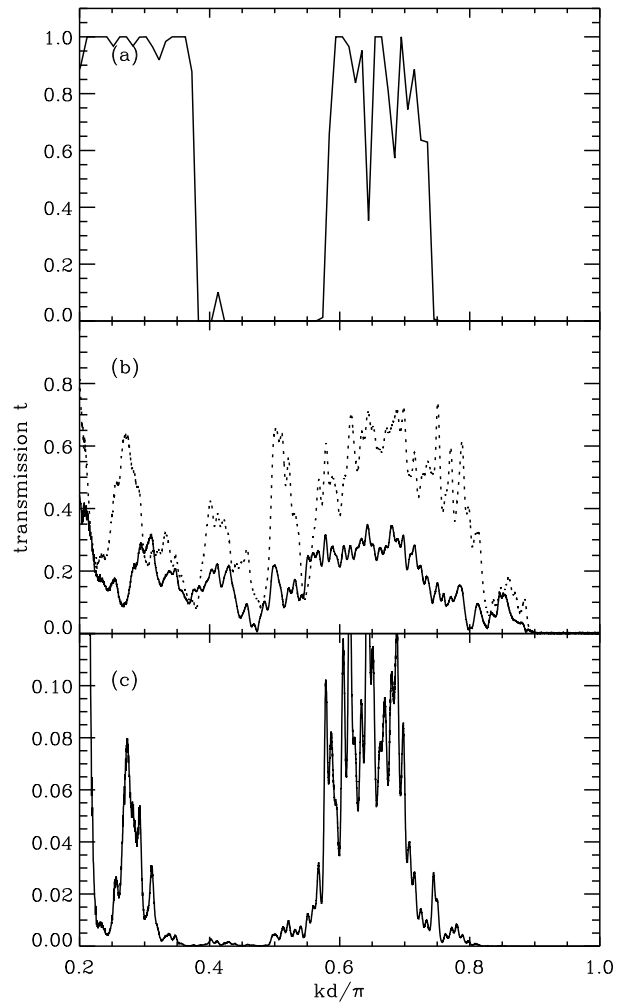,width=\FigWidth} \narrowtext
\caption{\label{fig:band2} Same as Fig.~\ref{fig:band1} but for
the complementary potential, see in the text.}
\end{figure}

A sequence of scattering strengths $\{\epsilon_n\}$ of the length
$N=10^4$ was generated by calculating $\beta_m$ from
Eq.~(\ref{beta1}) with $\mu_1/\pi =0.2$ and $\mu_2/\pi =0.4$, and
substituting the result into Eq.~(\ref{alg}).
Fig.~\ref{fig:band1}(a) shows the resulting transmission for
$\sqrt{\langle\epsilon_n^2\rangle}=0.1$ and $\epsilon=-0.1$. This
value for $\epsilon$ was obtained from Eq. (\ref{shift}) by
adjusting the width of the allowed band to the experimental data.
Experimental data are shown in Figs. \ref{fig:band1}(b,c). The
mobility edges are clearly seen near the points $kd/\pi=0.38,
0.57, 0.76$ which are the roots of Eq. (\ref{shift}) with
$\mu/\pi=0.4, 0.6, 0.8$ at the r.h.s. Transmission spectrum ends
at the band edge, $kd/\pi=0.91$ for $\mu=\pi$. Data below
$kd/\pi=0.2$ are not shown because of strong absorption in the
waveguide at low frequencies.

We have also calculated
and measured the transmission through the potential which is complementary to the previos one.
Namely, it is
transparent in the regions $0.2<\mu<0.4$, $0.6<\mu<0.8$
and untransparent otherwise. For this case coefficients $\beta_m$ have
opposite sign, $\beta_0=\frac{2C_0}
{\pi}(\mu_1-\mu_2+\frac{\pi}{2})$,
and $C_0^2=\frac{\pi}{2(\mu_1-\mu_2+\pi/2)}$.
The results are shown in Fig. 3(a)
for the same parameters $\epsilon_0$,
$\epsilon$, and $N$ as above.
In this case again we got a reasonable agreement between
analytical, numerical, and experimental data for the positions of
the mobility edges.

For the experiment we used a segment of the length of $N=500$ from
each of mutually complementary random sequences in order to create
the appropriate scattering arrangements in the waveguide. The
$\epsilon_n$ were mapped into screw lengths by identifying the
minimum $\epsilon_n$ value with a length of 0~mm and the maximum
value with a length of 3~mm. The average screw length of 1.5~mm
determines the width of the Bloch bands \cite{kuh98b}. This
procedure ignores the phase shift due to the scatterers which
probably is responsible for the negative value of $\epsilon$ (see
above). Five measurements were performed with each random
sequence. In one measurement a transmission through a realization
of hundred scatterers have been measured. The dotted lines in
Fig.~\ref{fig:band1}b and \ref{fig:band2}b show the results of a
single measurement, the solid lines represent the results of
averaging over all five measurements. The expected transmission
pattern is already visible. But we can do even better. Since in
each measurement the complete $2 \times 2 $ scattering matrix
$S_n(n=1,..,5)$ has been obtained, the corresponding transfer
matrix $T_n$ is available as well. Then the total transfer matrix
$T$ is the matrix product,
$T=\prod_{n=1}^5 T_n$.
Thus it is possible to study the transmission through arbitrary long
sequences of scatterers with a set-up containing only 100 of them.
However, because of absorption this technique is limited in our case to a total
of about 1000 scatterers.
Figs.~\ref{fig:band1}c and \ref{fig:band2}c show such transmission spectra.
In both cases the expected
transparent and non-transparent regions are clearly reproduced.

\begin{figure}
\psfig{file=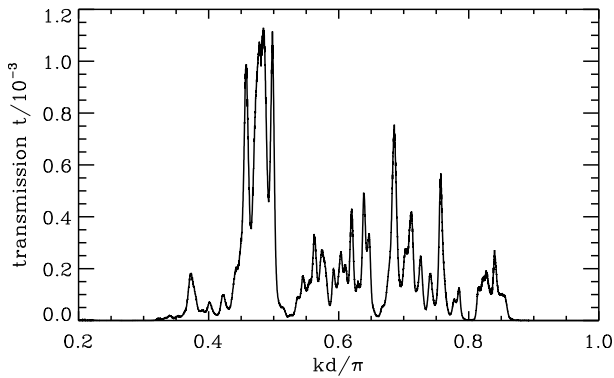,width=\FigWidth}\vspace*{1cm} \narrowtext
\caption{\label{fig:random} Microwave transmission through a
random arrangement of $N=500$ scatterers obtained by multiplying
the transfer matrices of five individual measurements.}
\end{figure}

As a check we studied in addition the transmission through
the uncorrelated random
sequence of 500 scatterers. The result is shown in Fig.~\ref{fig:random}.
Here the transmission is approximately 100 times smaller as
for the sequences with correlated disorder within the transparent region.

Finally we would like to mention that Eq. (\ref{base})
does not take into
account waveguide effects in the device. That is why one cannot expect
a complete similarity
in the behavior of the transmission coefficient obtained numerically with that
measured in the experiment.
Nevertheless, this equation allows one to calculate
a correlated potential  which being maped to the length of scatterers,
reproduces any prescribed structure of
transparent and untransparent frequency zones. This direct mapping turns out to be
rather succesful if only the lowest mode of the waveguide is excited.
For higher modes one might expect stronger influence of the
waveguide effects.

The idea to this cooperation was developed during a workshop
at the International Center for Sciences in Cuernavaca in November 1998.
Final discussions
took place during another workshop at the MPI for Complex Systems
in Dresden in May 1999. We thank the organizers of the workshops
for the invitations and the
institutions for their hospitality which made this work possible.
This work was supported by CONACyT (Mexico) Grants No. 26163-E and
No. 28626-E and by the DFG via the Sonderforschungsbereich 185.

\end{multicols}


\begin{thebibliography}{10}
\bibitem{ThanksSok}  We are thankful to V. Sokolov for the suggestion
concerning this algorithm.
\end{thebibliography}
\end{document}